\documentclass[showkeys,aps,twocolumn,showpacs,preprintnumbers,amsmath,amssymb,prl,letterpaper,floatfix,nofootinbib,superscriptaddress,]{revtex4-1}

\usepackage{amsmath,amssymb}
\usepackage[usenames,dvipsnames]{color}
\usepackage{graphicx}
\usepackage{subfigure}
\usepackage{soul}
\newcommand\bef{\begin{figure}}
\newcommand\eef[1]{\label{fg:#1}\end{figure}}
\newcommand\beq{\begin{equation}}
\newcommand\eeq[1]{\label{#1}\end{equation}}
\newcommand\bet{\begin{table}}
\newcommand\eet[1]{\label{tb:#1}\end{table}}
\newcommand\fgn[1]{Figure \ref{fg:#1}}

\newcommand\tbn[1]{Table \ref{tb:#1}}

\begin{document}
\title{Precise predictions of charmed-bottom hadrons from lattice QCD}

\author{Nilmani\ \surname{Mathur}}
\email{nilmani@theory.tifr.res.in}
\affiliation{Department of Theoretical Physics, Tata Institute of Fundamental
         Research,\\ Homi Bhabha Road, Mumbai 400005, India.}

\author{M.\ \surname{Padmanath}}
\email{Padmanath.M@physik.uni-regensburg.de}
\affiliation{Instit\"ut  f\"ur  Theoretische  Physik,  Universit\"at  Regensburg,
Universit\"atsstrase  31,  93053  Regensburg,  Germany.}

\author{Sourav\ \surname{Mondal}}
\affiliation{Department of Theoretical Physics, Tata Institute of Fundamental
         Research,\\ Homi Bhabha Road, Mumbai 400005, India.}


\begin{abstract}
We report the ground state masses of hadrons containing at least one charm 
and one bottom  quark using lattice quantum chromodynamics. These include mesons with spin ($J$)-parity ($P$) quantum
numbers ($J^{P}$):   $0^{-}$, $1^{-}$, $1^{+}$ and $0^{+}$ and the spin-1/2 and 3/2 baryons. Among these hadrons only the ground state of $0^{-}$ is known experimentally and therefore our predictions provide important information for the experimental discovery of all other hadrons with these quark contents.
\end{abstract}
\maketitle

Recently heavy hadron physics has attracted huge scientific interests
mainly due to the prospects of studying new physics beyond the
Standard Model at the intensity frontier
~\cite{Antonelli:2009ws,Bediaga:2012py,Bevan:2014iga,DAmico:2017mtc,Buttazzo:2017ixm}, and to study various newly discovered subatomic
particles to better understand the confining nature of strong
interactions ~\cite{Voloshin:2007dx,Hosaka:2016pey,Ali:2017jda,
  Karliner:2017qhf,Lebed:2016hpi,Esposito:2016noz,
  Guo:2017jvc}. From the perspective of newly found hadrons itself, a
large number of discoveries over the past decade ranging from usual
mesons \cite{Dobbs:2008ec,Ablikim:2010rc,Adachi:2011ji,Lees:2011zp,Bhardwaj:2013rmw,Ablikim:2015dlj,Chilikin:2017evr,Sirunyan:2018dff}, baryons~\cite{Aaij:2017ueg} along with their
excited states~\cite{Aaij:2017nav,Padmanath:2017lng,Aaij:2017vbw,Aaij:2018yqz}, to new exotic particles like
tetraquarks~\cite{Choi:2007wga, Aaij:2014jqa, Belle:2011aa} and
pentaquarks~\cite{Aaij:2015tga}, as well as hadrons whose structures
are still elusive~\cite{Choi:2003ue, Aubert:2005rm, Yuan:2007sj,
  Voloshin:2007dx, Hosaka:2016pey, Ali:2017jda, Aaij:2013zoa}, have
proliferated interests in the study of heavy hadrons. Furthermore, it
is envisaged that the large data already collected or to be obtained
at different laboratories, particularly at LHCb and Belle II,
will further unravel many other hadrons. One variety of such theorized but as yet essentially unobserved (except one)
subatomic particles are hadrons made of at least a charm and a bottom
quarks, the charmed-bottom ($bc$) hadrons.

Investigations of such hadrons are highly appealing, as they provide a unique laboratory to explore the heavy quark dynamics at multiple scales:~$1/m_b$, $1/m_c$ %($m_Q$: mass of heavy quark)
and $1/\Lambda_{QCD}$.
%, and hence may reveal more information on strong interactions which may be concealed in other hadrons.
Decay constants and form factors of $bc$ mesons are still unknown but are quite important because of their relevance to investigate physics beyond the standard model, particularly in view of the recent measurement of $R(J/\psi)$~\cite{Aaij:2017tyk}.
The information on spin
splittings and decay constants can shed light
on their structures and help us to understand the nature of strong interactions at multiple scales. Moreover, $bc$ baryon decays can aid in studying $b \rightarrow c$ transition and  $|V_{cb}|$ element of the CKM matrix.

However, to date the discovery of these hadrons is limited to only two observations:   $B_c (0^-)$ with mass 6275(1) MeV~\cite{Patrignani:2016xqp} and $B_c(2S) (0^-)$ at 6842(6) MeV~\cite{Aad:2014laa} while the latter has not yet been confirmed~\cite{Aaij:2017lpg}. On the other hand, LHC being an efficient factory for producing $bc$ hadrons~\cite{Berezhnoy:1997fp, Gouz:2002kk}, one would envisage for their discovery and study their decays in near future. Precise theoretical predictions related to the energy spectra and decay of these hadrons are thus utmost essential to guide their discovery. % of these states. 

In fact various model calculations exist in literatures on $bc$ mesons 
\cite{Kwong:1990am, Eichten:1994gt, Kiselev:1994rc, Gupta:1995ps, Fulcher:1998ka, Ebert:2002pp, Godfrey:2004ya} and baryons 
\cite{Roberts:2007ni, Karliner:2008sv, Kiselev:2001fw, Ebert:2011kk, Karliner:2014gca, Shah:2016vmd, Karliner:2018hos}. However, those predictions vary widely, e.g. 1S-hyperfine splitting in $B_c(\bar{b}c)$ mesons 
spread over a range of 40-90 MeV~\cite{Kwong:1990am, Eichten:1994gt, Kiselev:1994rc, Gupta:1995ps, Fulcher:1998ka, Ebert:2002pp, Godfrey:2004ya}.  The predictions on $bc$ baryons and excited states are even more scattered. Naturally first principle
calculations using lattice QCD % with controlled systematics
%, which have already proved to be successful 
%in predicting the masses of low lying \cite{Aubin:2004wf, Durr:2008zz, Aoki:2008sm, Bazavov:2009bb, Dowdall:2012ab, Borsanyi:2014jba} and excited hadrons \cite{Dudek:2011tt,Edwards:2012fx,Padmanath:2013zfa},
 are quite
essential to study these hadrons. However, unlike quarkonia, lattice study of $bc$ hadrons are confined only 
to a few calculations~\cite{Gregory:2009hq, Dowdall:2012ab, Wurtz:2015mqa, Brown:2014ena, Mathur:2002ce}. 
In this work we carry out a detailed lattice calculation of the ground state energy spectra of all low lying $bc$ hadrons (showed in \tbn{bc_hadrons}) with very good control over 
systematics and predict their masses most precisely to this date.
\bet[h]
\centering
\begin{tabular}{c|ccccc}
\hline
{Mesons($\bar{q_1}q_2$)}          & \multicolumn{5}{c}{Baryons ($[q_1q_2q_3] (J^P)$)} \\
\hline
                           &   $J^{P} \equiv$          $1/2^+$ &&                      $1/2^+$ &&                  $3/2^+$ \\\hline
$B_c(\bar{b}c)(0^{-})$     &     $\Xi_{cb}[cbu]$ &&     $\Xi_{cb}^{\prime}[cbu]$ &&      $\Xi_{cb}^{*}[cbu]$ \\
$B_c^{*}(\bar{b}c)(1^{-})$ &  $\Omega_{cb}[cbs]$ &&  $\Omega_{cb}^{\prime}[cbs]$ &&   $\Omega_{cb}^{*}[cbs]$ \\
$B_c(\bar{b}c)(0^{+})$     & $\Omega_{ccb}[ccb]$ &&  &&  $\Omega_{ccb}^{*}[ccb]$ \\
$B_c(\bar{b}c)(1^{+})$     & $\Omega_{cbb}[bbc]$ &&  &&  $\Omega_{cbb}^{*}[bbc]$ \\
\hline
\end{tabular}
\caption{List of $bc$ hadrons that we study in this work. Quantum numbers ($J^P$) along with the valence quark contents are also mentioned.}
\eet{bc_hadrons}

Lattice QCD studies are subject to various lattice artefacts. Of these the most relevant 
one in a study of heavy hadrons is the discretization error.
% due to large masses of heavy quarks.
 It is thus essential to take a controlled continuum extrapolation of the results from finite lattice spacings. To that goal we obtain results at three lattice spacings: $a\sim$ 0.12, 0.09 and 0.06 {\it{fm}}, and then are able to perform such extrapolations. Below we elaborate our numerical procedure.

\noindent{\bf{Numerical details:}} \,\,
\noindent{\bf{A. Lattice ensembles: }}
We use three dynamical 2+1+1 flavours ($u/d,s,c$) lattice ensembles generated by the MILC collaboration~\cite{Bazavov:2012xda} with HISQ fermion action~\cite{Follana:2006rc}. The lattices are with sizes  $24^3 \times 64$, $32^3 \times
96$ and $48^3 \times 96$ at gauge couplings $10/g^2 = 6.00,
6.30$ and $6.72$, respectively~\cite{Bazavov:2012xda}.
%The details of these gauge
%configurations, which are currently being extensively used by MILC-Fermilab-HPQCD collaborations, are summarized in Ref.~\cite{Bazavov:2012xda}.
The measured lattice
spacings, obtained from $r_1$ parameter, for the set of ensembles being used here are 0.1207(11) 0.0888(8) and 0.0582(5) {\it{fm}}, respectively~\cite{Bazavov:2012xda}.

\noindent{\bf{B. Quark actions:}}
For valence quark propagators, from light to
charm quarks, we use the overlap action
which has exact chiral symmetry at finite lattice spacings~\cite{Neuberger:1997fp, Neuberger:1998wv, Luscher:1998pqa} and no $\mathcal{O}(ma)$ error.
A wall source is utilized 
for calculating quark propagators. % for light to charm quarks.

%Using multimass algorithm of overlap action~\cite{Edwards:1998wx, Chen:2003im}, we are able simulate light to charm quarks
%in a single lattice formalism with little computational overhead.
%A wall source is utilized as smearing function for calculating the quark propagators on Coulomb gauge-fixed lattices.

For bottom quark we utilize a non-relativistic QCD (NRQCD)
formulation \cite{Lepage:1992tx} in which
we incorporate all terms up to % $1/M_{0}^2$ and
the leading
term of the order of $1/M_{0}^3$, where $M_{0} = am_b $ is the
bottom mass~\cite{Lewis:2008fu,
  Mathur:2016hsm}.
%bare
%mass of the bottom quark~\cite{Lewis:2008fu,
%Mathur:2016hsm}.
  This Hamiltonian is improved by including
spin-independent terms through $\mathcal{O}(\alpha_sv^4)$ with
non-perturbatively tuned improvement coefficients\cite{Dowdall:2011wh}.
%, as estimated in
%Ref. \cite{Dowdall:2011wh} for these ensembles. Such an action has been utilized by HPQCD collaboration over many years and found to be very effective
%for bottom quark physics \cite{Gregory:2009hq, Dowdall:2011wh, Dowdall:2012ab, Na:2012kp, Dowdall:2013tga}.
For the coarser two ensembles, we study the spectrum using {\it ``improved''}
coefficients as well as tree level
coefficients ({\it ``unimproved''}).
%For the finer lattice, since the improved coefficients have not been calculated yet we study only with tree level coefficients.

\noindent {\textbf{C. Quark mass tuning:} Following the Fermilab prescription for heavy quarks~\cite{ElKhadra:1996mp} we tune the heavy quark masses by equating the spin-averaged kinetic mass of the $1S$ quarkonia states 
  ($\bar{M}_{kin}(1S) = {3\over 4} M_{kin}(1^-) + {1\over 4} M_{kin}(0^-)$) to their
%respective
  physical values~\cite{Basak:2013oya, Mathur:2016hsm}. A momentum 
induced wall-source, which is found to be very efficient compared to point or smeared sources \cite{Basak:2012py}, is
utilized to obtain kinetic masses precisely.
%with 
%significantly little statistics.
%obtain energy values from the correlators with finite momenta and helps to obtain kinetic masses precisely with 
%significantly little statistics.
%Details of the charm and bottom quark mass tuning are described in Ref.~\cite{Basak:2013oya} 
%and Ref.~\cite{Mathur:2016hsm}, respectively.
The tuned bare charm quark masses  are found to be 0.528, 0.427 and 0.290 on coarse to fine lattices respectively, which also satisfy $m_ca << 1$, a necessary condition
%for using heavy relativistic quark
for reducing discretization effects. We tune strange quark mass, following Ref.~\cite{Chakraborty:2014aca}.
%, by equating the unphysical 
%$\bar{s} s$ pseudoscalar mass to 688.5 MeV~\cite{Basak:2012py,Basak:2013oya}. 

\noindent\textbf{D. Hadron interpolators:} For mesons, we utilize the local meson interpolators ($\bar b \Gamma c$),
where $\Gamma$, corresponding to different spin($J$) and parity($P$) quantum numbers, $J^P$, are : $\gamma_5 (0^-)$, $\gamma_i (1^-)$, $I (0^+)$
and $\gamma_5\gamma_i (1^+)$. We work with the assumption that the extracted ground state with 
$\gamma_5\gamma_i$ is $1^+$ and is unaffected by a possible nearby 
$2^+$ level \cite{Dowdall:2012ab}. 
For baryons, we utilize the conventional interpolators given by $P^{+}[(q_{1}^{T} C\Gamma q_2) q_3]$
as discussed in detail in Refs.~\cite{Mathur:2002ce, Lewis:2001iz, Brown:2014ena}. For spin-1/2 $\Xi_{cb}$ and $\Omega_{cb}$,
$\Gamma = \gamma_5$, whereas for spin-1/2 $\Xi_{cb}^{\prime}$, $\Omega_{cb}^{\prime}$ and
spin-3/2 $\Xi_{cb}^*$, $\Omega_{cb}^*$ we use $\Gamma = \gamma_i (i=1,2,3)$ with appropriate spin projections. A subtlety in the $\Xi_{cb}'$ correlators
is the possible admixture of $\Xi_{cb}$ baryons. However, the use of wall source help us to clearly distinguish these two correlators which suggest that these two correlators coupled to two distinct states with no significant admixture. In the heavy quark limit, the total spin of the $bc$ diquark becomes a good quantum number, and thus the mixing is heavily suppressed. An agreement between our results on these baryons with those obtained in Ref.~\cite{Brown:2014ena} also justify this observation. Below we elaborate our results.

\noindent {\textbf{Results:} To cancel out bare quark mass term which enters additively into the NRQCD Hamiltonian we calculate the mass differences between energy levels, rather than masses directly. To obtain the mass of a hadron ($M^c_{H}$) we first 
calculate subtracted masses on the lattice as
\beq
\Delta M_{H} = [M^{L}_{H} - n_b\overline{1S}_b/2 - n_c\overline{1S}_c/2]a^{-1},
\eeq{submass}
where $\overline{1S}_b$ and $\overline{1S}_c$ are the lattice calculated spin average $\overline{1S}$ bottomonia and 
charmonia masses respectively, whereas $n_b$ and $n_c$ are the number of $b$ and $c$ valence quarks 
in the hadron.
After calculating 
%At each lattice spacing we calculate
 this subtracted mass we perform the 
continuum extrapolation to get its continuum value $\Delta M^{c}_{H}$. Finally the physical result is obtained by
adding the physical values of spin average masses to $\Delta M^{c}_{H}$ as 
\begin{equation}
M^c_{H} = \Delta M^{c}_{H} + n_b(\overline{1S}_b)_{phys}/2 + n_c(\overline{1S}_c)_{phys}/2.
\end{equation}
Since the $B_c(0^-)$ mass is known experimentally we also utilize a dimensionless ratio,
\begin{equation}
	R_{H} =  {{M^{L}_{H} - n_b\overline{1S}_b}/2\over{M^{L}_{B_c(0^{-})} - n_b\overline{1S}_b}/2},
\end{equation}
which is then extrapolated to the continuum limit ($R^{c}_{H}$) and the final hadron mass is obtained from
\begin{equation}
M^c_{H} = R^{c}_{H} \times (M_{B_c(0^{-})} - n_b\overline{1S}_b/2)_{phys} + n_b(\overline{1S}_b)_{phys}/2.
\end{equation}
These procedures of utilizing dimensionless ratios as well as mass differences for the continuum extrapolations substantially reduce the systematic errors arising from mass tuning as well as for the terms which enter masses additively. 
We use both equations (2) and (4) and found consistent results and added the difference in systematics. Below we discuss results for $bc$ mesons and baryons.
\bef[tbh]
\vspace*{-0.29in}
\centering
\includegraphics*[height=2.0in,width=3.3in]{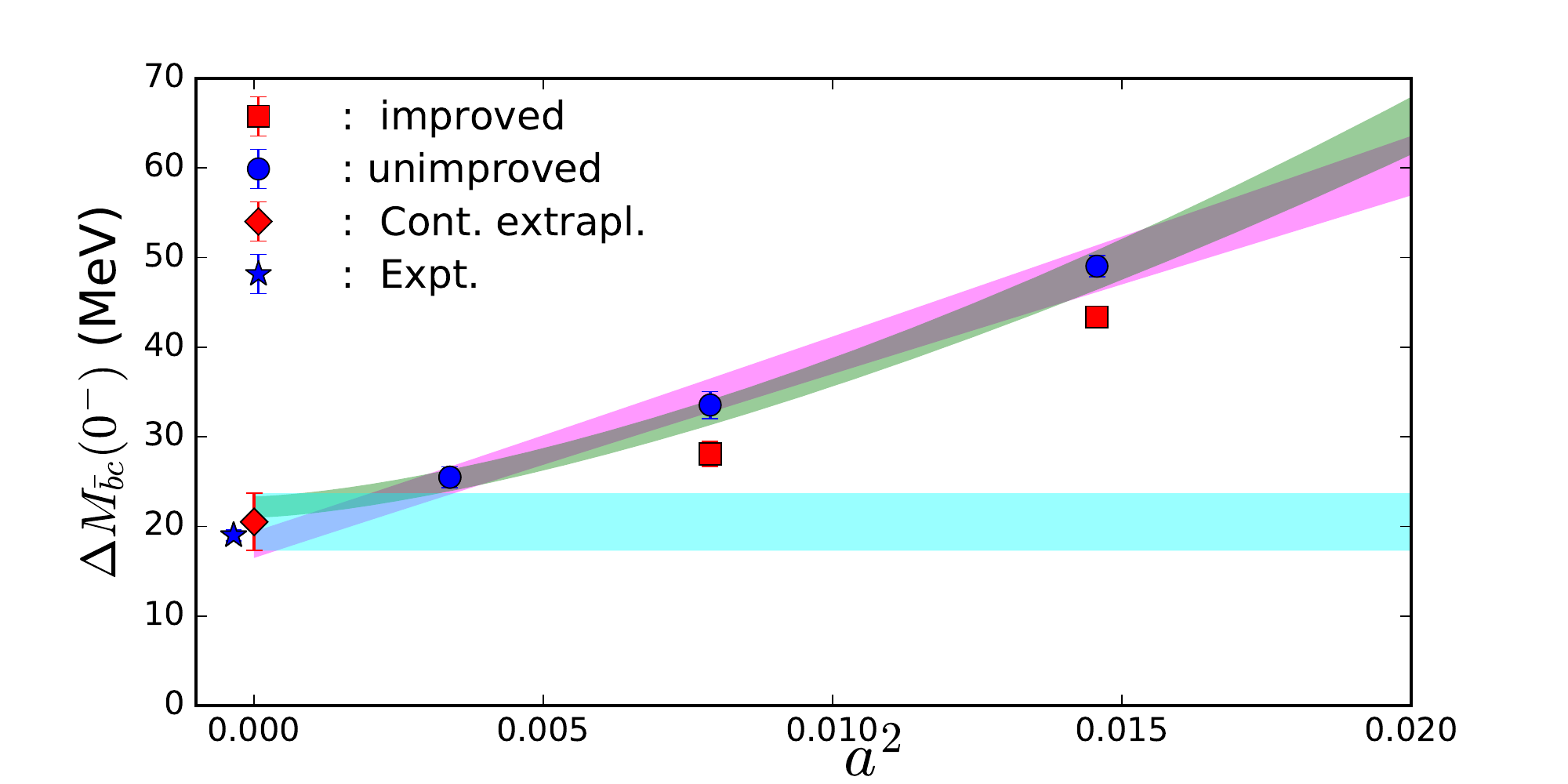}
\vspace*{-0.09in}
\caption{Ground state mass of the $B_c(0^{-})$ meson at three lattice spacings are plotted in terms of energy splittings from the spin-average mass (Eq(1)). Continuum extrapolated and experimental values are also shown.}
\eef{fig_bc_spec}
\bef[tbh!]
\vspace*{-0.29in}
\centering
\hspace*{-0.1in}
\includegraphics*[height=3.7in,width=3.7in]{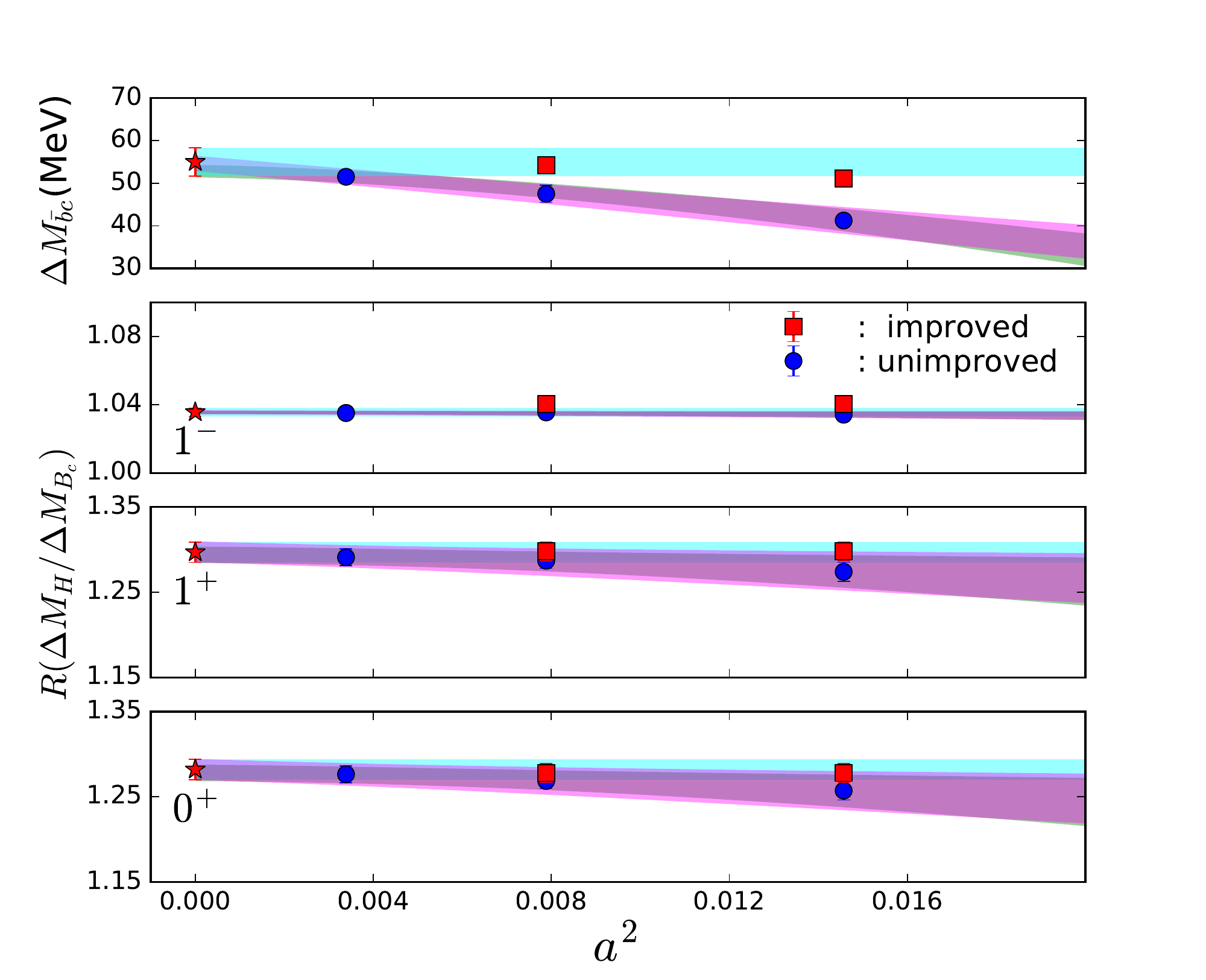}
\caption{(Top) Hyperfine splitting in $B_c$ mesons at three different lattice spacings and at the continuum limit. (Bottom three) Ratios (as defined in Eq.(5)) of subtracted masses of $1^{-}$, $1^{+}$ and $0^{+}$ mesons to that of $0^{-}$ $B_c$ meson are plotted at three lattice spacings and at the continuum limit.}
\vspace*{-0.2in}
\eef{fig_bc_mesons}

\noindent {\textbf{Mesons:} In~\fgn{fig_bc_spec} we plot the subtracted mass ($\Delta M_{H}$), 
as defined in Eq. (1), for $B_c(0^{-})$ as a function of lattice spacings ($a$). Blue circles 
represent {\it unimproved} and red squares represent {\it improved} results. We extrapolate {\it unimproved} results
using fit forms $Q^f=A+a^2B$ as well as $C^f=A+a^3B$. Two bands corresponds to one sigma error 
for these fittings (purple: $Q^f$, green: $C^f$).  The extrapolated result and the experimental 
value are shown by red and blue stars respectively.
As expected the {\it improved} results are 
closer to the continuum limit (horizontal cyan bands show the proximity of the {\it improved}
results from the continuum result).
%than that of {\it unimproved} ones.
%With only two {\it improved}
%points no extrapolation is performed of their own but we fit these together with the unimproved point of the fine lattice and that aids in determining the central value and the extent of discretization error.
%We also draw a horizontal cyan band, to show the proximity of the {\it improved} results from the continuum result.
 To see the consistency in fits we also use a constrained fit with both forms together by loosely constraining $A$ values from previous fits and difference in fitted parameters are included in discretization error.
As in ~\fgn{fig_bc_spec}, throughout we follow 
the same conventions for symbols and color coding. In~\fgn{fig_bc_mesons} (top), we plot 
the hyperfine splitting of 1S $B_c$ mesons. After the continuum extrapolation we obtain $B^{*}_c(1^-) - B_c(0^-) = 55(3)$ MeV which is consistent 
with previous lattice calculations~\cite{Dowdall:2012ab, Wurtz:2015mqa} but more precise. In the bottom figure we show 
the subtracted ratios (Eq.(3)) and continuum extrapolations for the ground states 
of $1^-$, $1^+$ and $0^+$ $B_c$  mesons. Taking the experimental values for
$B_c(0^{-})$ and $\overline{1S}$ quarkonia~\cite{Patrignani:2016xqp} masses, we obtain the ground state masses for these
mesons and tabulate those in \tbn{bcbres}.

\noindent {\textbf{Baryons:} We first discuss the $\Xi_{cb}$ baryons. Presence 
of a valence light quark in $\Xi_{cb}$ demands a chiral extrapolation. Use of multimass algorithm allows to simulate a range 
of pion masses.
%and for $24^3 \times 64$ lattice we even calculate quark propagators close to the physical one.
 In \fgn{fig_bcu} (top),
%for this lattice
 we plot $\Xi_{cb}$ masses at various pion masses which clearly show a quadratic 
variation starting from the physical pion mass to $\sim600$ MeV. We 
thus use a chiral extrapolation of the form $A+m_{\pi}^2B$.
Within the limit of acceptable $\chi^2$/dof,  variations in chiral extrapolation forms, as in Ref.~\cite{Brown:2014ena}, 
do not change the final value. The same procedure is repeated for $\Xi_{cb}^{\prime}$ 
and $\Xi_{cb}^{*}$ at three lattices. These chiral extrapolated values are then used 
to calculate the subtracted masses and are plotted in the bottom part of \fgn{fig_bcu}. 
These subtracted masses are then extrapolated to the continuum limit 
to get % final values for
the ground state masses of these baryons and are tabulated in
\tbn{bcbres}. In Figs. \ref{fg:fig_bcs_and_all}, we 
show lattice extracted $\Delta M_H$ and the continuum extrapolations for different $\Omega$ baryons 
with flavor content $bcs$, $bcc$ and $bbc$, respectively.
%No chiral extrapolation is involved for these baryons.
Continuum extrapolated 
results are shown by stars in each figure and are listed in \tbn{bcbres}.}

\bef[h!]
\vspace*{-0.19in}
\centering
\includegraphics*[height=3.8in,width=3.7in]{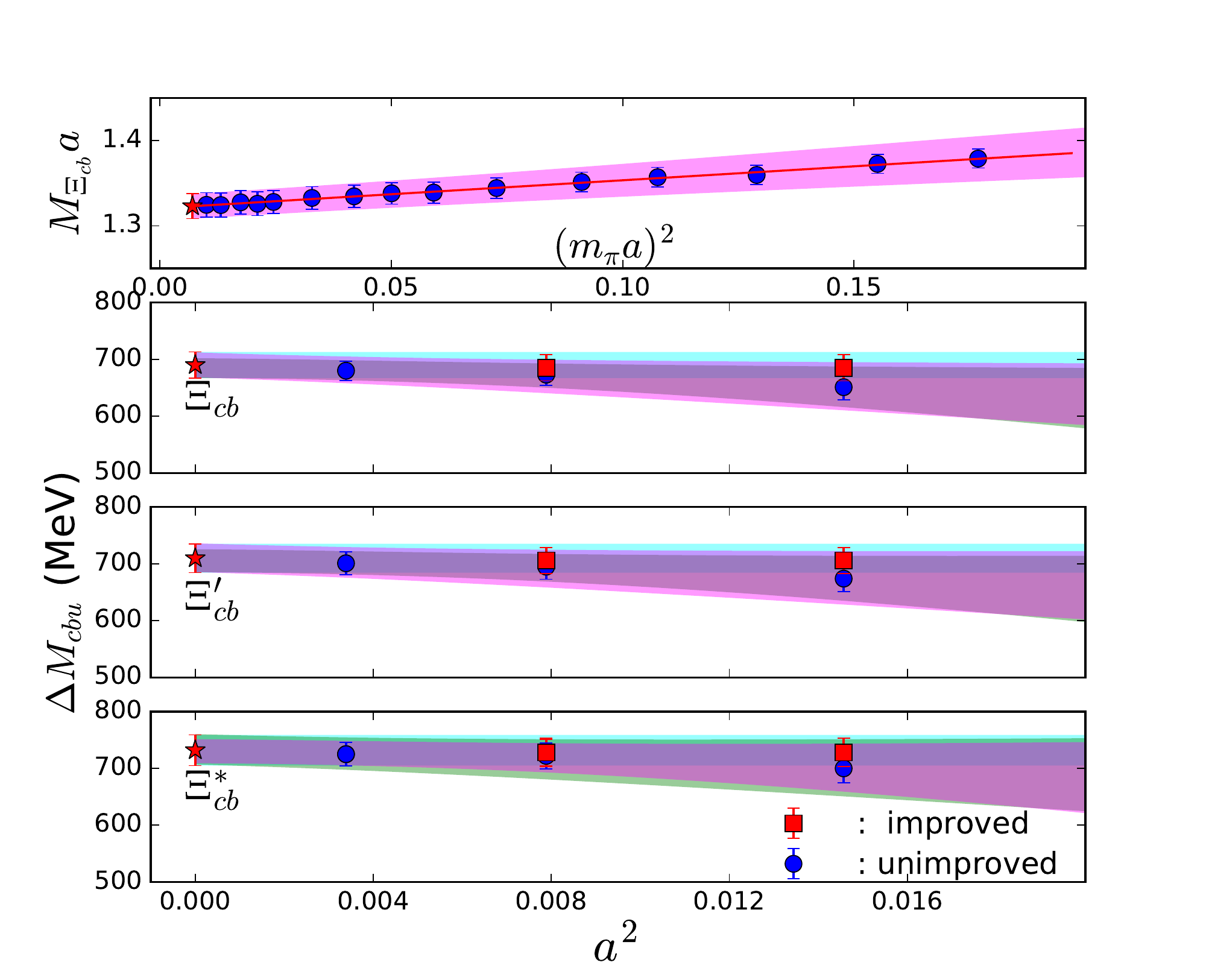}
%\vspace*{-0.05in}
\caption{(Top) $\Xi_{cb}$ masses as a function of $m_{\pi}^2$. Chiral extrapolation is performed with a linear term in $m_{\pi}^2$. (Bottom three)  Chiral extrapolated (as shown above) subtracted masses (as defined in Eq.(1)) of $\Xi_{cb}$, $\Xi^{\prime}_{cb}$ and $\Xi^{*}_{cb}$ baryons at three lattice spacings and at the continuum limit.}
\vspace*{-0.09in}
\eef{fig_bcu}
\bet[h]
\centering
\begin{tabular}{lc|clc }
\hline
Hadrons           &&& Lattice      & Experiment \\
\hline              
$B_c(0^{-})$     &&& 6276(3)(6)   & 6274.9(8) \\
$B_c^{*}(1^{-})$ &&& 6331(4)(6)   & ?         \\
$B_c(0^{+})$     &&& 6712(18)(7) & ?         \\
$B_c(1^{+})$     &&& 6736(17)(7) & ?         \\
\hline                             
$\Xi_{cb}(cbu)(1/2^+)$              &&& 6945(22)(14) &? \\
$\Xi_{cb}^{\prime}(cbu)(1/2^+)$     &&& 6966(23)(14)&?  \\
$\Xi^{*}_{cb}(cbu)(3/2^+)$          &&& 6989(24)(14)&?  \\
$\Omega_{cb}(cbs)(1/2^+)$           &&& 6994(15)(13)&?  \\
$\Omega_{cb}^{\prime}(cbs)(1/2^+)$  &&& 7045(16)(13)&?  \\
$\Omega^{*}_{cb}(cbs)(3/2^+)$       &&& 7056(17)(13)&?  \\
$\Omega_{ccb}(1/2^+)$               &&& 8005(6)(11)  &? \\
$\Omega^{*}_{ccb}(3/2^+)$           &&& 8026(7)(11)   &?\\
$\Omega_{cbb}(1/2^+)$               &&& 11194(5)(12)  &?\\
$\Omega^{*}_{cbb}(3/2^+)$           &&& 11211(6)(12)  &?\\
\hline
\end{tabular}
\caption{Ground state masses of $B_c$ mesons and baryons as predicted in this work. Statistical and systematic errors are shown inside two parentheses respectively.}
\eet{bcbres}
\bef[h]
\centering
\vspace*{-0.22in}
\hspace*{-0.15in}
\includegraphics*[height=4in,width=3.9in]{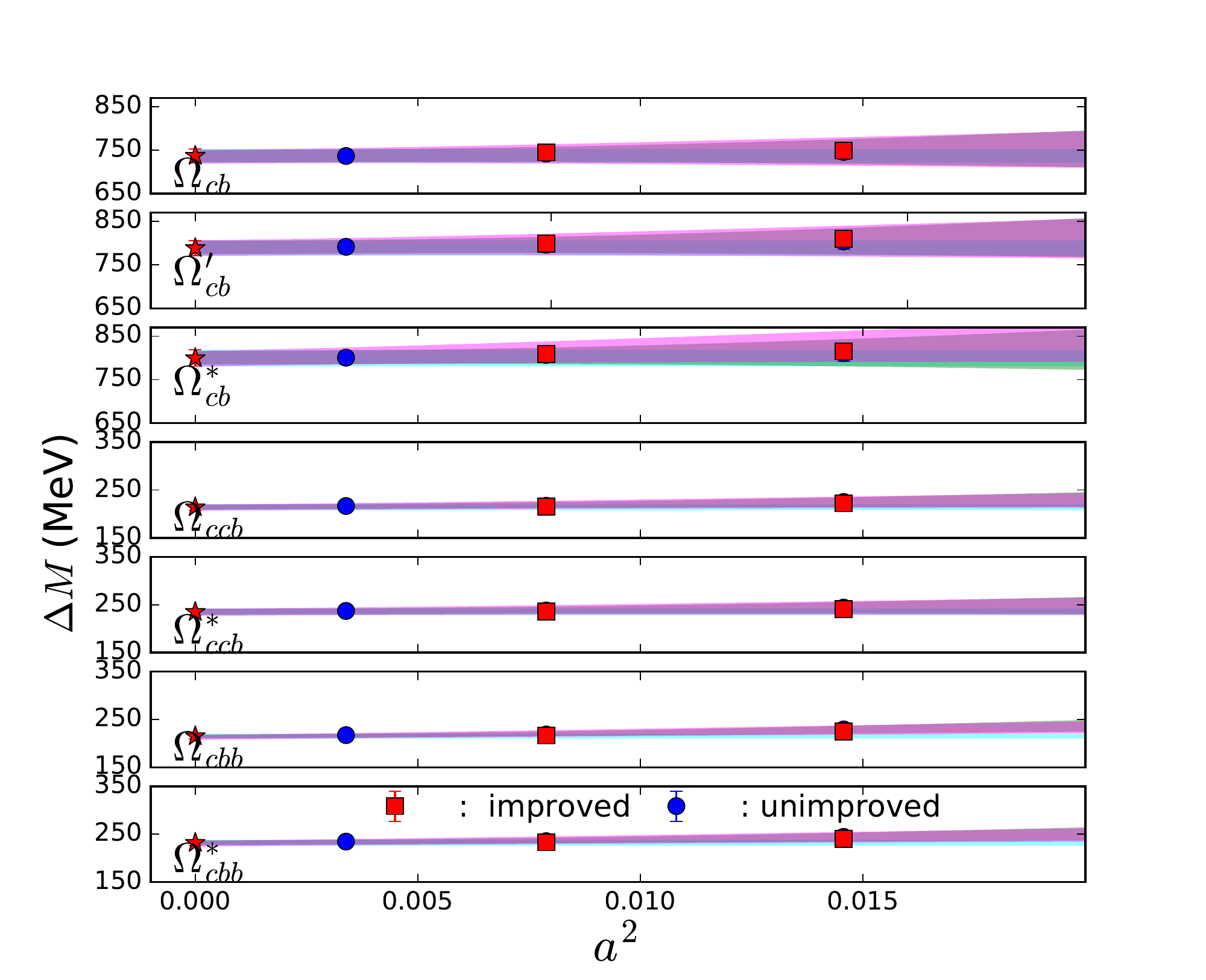}
\vspace*{-0.2in}
\caption{Subtracted masses, as defined in Eq. (1), of $\Omega_{cb}$, $\Omega^{\prime}_{cb}$, $\Omega^{*}_{cb}$ (top 3), $\Omega_{ccb}$, $\Omega^{*}_{ccb}$ (middle 2), and $\Omega_{cbb}$, $\Omega^{*}_{cbb}$ (bottom 2) baryons at three lattice spacings and at the continuum limit.}
\vspace*{-0.09in}
\eef{fig_bcs_and_all}

%%%%%%%%%%%%%%%%%%%%%
%\bef[tbh]
%\centering
%%\vspace*{-0.2in}
%\includegraphics*[scale=0.37]{bcc.pdf}
%%\vspace*{-0.08in}
%\caption{Same as Fig. (4) but for $\Omega_{ccb}$ and $\Omega^{*}_{ccb}$ baryons.}
%\eef{fig_bcc}
%\bef[tbh]
%\centering
%%\vspace*{-0.2in}
%\includegraphics*[scale=0.37]{bbc.pdf}
%%\vspace*{-0.08in}
%\caption{Same as Fig. (4) but for $\Omega_{cbb}$ and $\Omega^{*}_{cbb}$ baryons.}
%\eef{fig_bbc}
%%%%%%%%%%%%%%%%%%%
\noindent \textbf{Error estimation:} Below we address the estimation of various errors related to this work.

\noindent \textit{Statistical}:  The use of wall source reduces the statistical 
errors substantially and facilitates wide and stable fit ranges even for baryons. We find that 
the statistical error is always below percent level and is maximum for the $\Xi_{cb}$ baryons which is about 0.4\%.

\noindent \textit{Discretization}:
Adaptation of overlap fermions ensures no $\mathcal{O}(ma)$ error 
for light to charm quarks. The value of $ma$ for charm quarks (0.528, 0.427 and 0.290 on three lattices) are rather small compared 
to unity and hence implies smaller error from higher orders in $ma$. The utilization of energy splittings and ratios %, for the continuum extrapolations
also ensure reduced systematics. This is clearly reflected in our estimates~\cite{Mathur:2016hsm} for quarkonia hyperfine splittings 
($\Delta E_{hfs}^{1S,\bar cc} = 115(2)(3)$ MeV and $\Delta E_{hfs}^{1S,\bar bb} = 63(3)(3)$ MeV). %, which are very much in agreement with their experimental values.
    These splittings are known to be quite susceptible to this error and an excellent agreement between our and experimental values assures good control over discretization and hence a reliable estimation of masses of other heavy hadrons. Different fitting methods, quadratic, cubic in lattice spacing as well as both together in constrained fits, help to access possible discretization effects in continuum extrapolations. The largest discretization error is found to be for $\Xi_{cb}$ baryons which is about 6-7 MeV.

    \noindent \textit{Scale setting}: We independently calculate lattice spacings from $\Omega_{sss}$ baryon mass and found those to be 
    %    obtained those as 0.1192(14), 0.0877(10) and 0.0582(5) {\it{fm}}, which are
    consistent with the values measured by MILC collaboration~\cite{Bazavov:2012xda}. %With these estimates of lattice spacings,
  The largest error in mass splittings due this scale uncertainty are within 6 MeV. 
  
  \noindent \textit{Finite size}:
The lattice volumes in this study is about $3 fm$. Furthermore, the hadrons considered are quite heavy and are mostly stable to strong decays (there is no negative parity baryons). $\Xi_{cb}$ baryons, only hadrons with 
valence light quark content, are found to have a perfect quadratic light quark 
mass dependence even towards the chiral limit indicating no observable finite 
size effects in them. Conservatively, we include a maximum uncertainty of a 
few MeV due to finite size effects, as estimated in Ref.~\cite{Brown:2014ena} 
on similar lattice volume. 

%The lattice volume in this study is about $3 fm$ and we believe it is large enough for all the hadrons studied here. These hadrons are quite heavy and stable to strong decays, and hence the finite volume effects are expected to be within a few MeV as estimated in Ref.~\cite{Brown:2014ena} on similar lattice volume.

\noindent \textit{Chiral extrapolation}:    
In this study only $\Xi_{cb}$ baryons are subjected to this error.
Due to the use of multimass algorithm we could calculate these baryons at a large number of pion masses, as shown in \fgn{fig_bcu}, which help to perform extrapolations to the physical limit in a controlled and reliable way. Our results are found to be quite robust with respect to different chiral extrapolation forms.

\noindent \textit{NRQCD errors}: Since we have included terms up to $\alpha_sv^4$, higher order terms, such as spin dependent as well as spin independent terms ($\alpha_s^2v^4$ and $\alpha_sv^6$) will contribute to the systematics. For $bc$ mesons, these errors are 4 MeV as estimated in Ref.~\cite{Dowdall:2012ab} on similar lattices. As in Ref. \cite{Brown:2014ena}, we also estimate these errors to be 5, 5 and 6 MeV for $bcq$, $bcc$ and $bbc$ baryons, respectively.

\noindent \textit{Other errors}: Errors due to quark mass tuning are expected to be negligible in these results, considering the precision and rigor
that enter into heavy quark mass tuning procedure. Use of wall source efficiently damps out excited state contamination
providing long plateau in the effective mass at sufficiently
large times indicating very good ground state saturation.
Hence, any related uncertainties in our calculation are also negligible
in comparison with any other errors. In a previous study we also found that the mixed action effects, which would vanish at the continuum limit, to be small~\cite{Basak:2014kma} within this lattice set up. As discussed in Ref.~\cite{McNeile:2012qf, Dowdall:2012ab, Chakraborty:2014aca} for similar lattices, the effect due to unphysical sea quark masses could be less than a percent level. Other errors due to electromagnetism, isospin breaking and the absence of dynamical bottom quarks are expected to be within 2-4 MeV~\cite{Dowdall:2012ab}.

As examples, following are the systematic error budget (in MeV) for $(B_c(0^{-}),~\Omega_{cbb})$: discretization (3, 5), scale setting (2, 6), 
NRQCD errors (4, 7), finite volume (0, 2) and other sources (3, 5)  which when are added in quadrature lead to systematic errors as $\sim (6, 12)$ MeV.

\bef[tbh]
\centering
%\vspace*{0.12in}
%\includegraphics*[scale=0.45]{Fig5.pdf}
\includegraphics*[height=1.8in,width=3.2in]{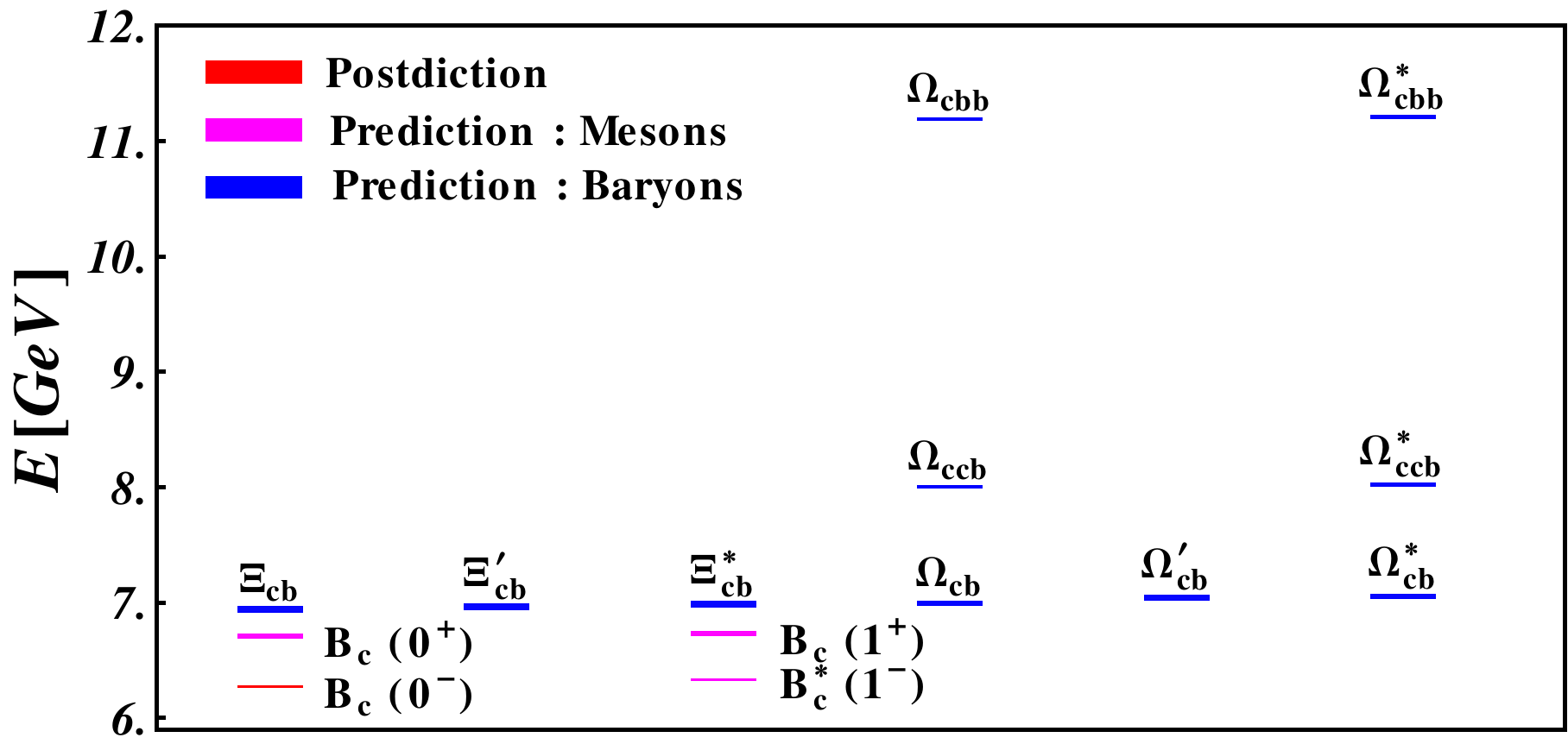}
%\vspace*{-0.05in}
\caption{Ground state masses of all $bc$ baryons as predicted in this work.}
\eef{fig_summary}

\noindent {\textbf{Summary: }
In this Letter, we present precise predictions of the ground state masses of $bc$ hadrons using lattice QCD simulations with very good control over systematics. These hadrons have not been discovered yet 
and considering the recent interests on them, particularly for their relevance to the physics beyond the Standard Model, these predictions provide important information for their future discovery. Our results are based on three different lattice spacings, finest one being 0.0582 fermi, which help us to
obtain precise results at the continuum 
limit. The overlap fermions, which have
%exact chiral symmetry at finite lattice spacings and
no ${\mathcal{O}}(ma)$ errors, are used for the light, strange as well as for the charm quarks. For the bottom quark, we use a non-relativistic formulation with non-perturbatively tuned coefficients 
with terms up to $\mathcal{O}(\alpha_sv^4)$. Utilization of a wall source helps to extract these masses unambiguously keeping the statistical error 
below percent level. Use of mass 
differences as well as ratios, in which the extent of discretization effects are significantly lesser for the continuum extrapolation, enables us to predict the masses % of these states more
precisely. We have 
 also addressed other possible systematic errors in detail, which when added in quadrature are  found to be smaller than the statistical error in most cases.
 Our final results for the ground state masses of all $bc$ hadrons are tabulated in Table II and also showed in ~\fgn{fig_summary}.}

\section*{Acknowledgements}
%{\bf{Acknowledgements:}}
We thank our colleagues within the ILGTI collaboration.
We are thankful to the MILC collaboration 
and in particular to S. Gottlieb for providing us with the HISQ lattices.  We would like to thank R. Lewis for helping with NRQCD code. We would also like to thank an unknown referee who has provided valuable comments on the mixing between $\Xi_{cb}$ and $\Xi_{cb}^{\prime}$.
Computations are carried out on the Cray-XC30 of ILGTI, TIFR, 
and on the Gaggle/Pride clusters of the Department of Theoretical Physics,
TIFR. N. M. would like to thank Christine Davies for discussions and also to Ajay Salve and P. M. Kulkarni for computational supports. M. P. acknowledges support from EU under grant no. MSCA-IF-EF-ST-744659 (XQCDBaryons) and 
the Deutsche Forschungsgemeinschaft under Grant No.SFB/TRR 55.

\bibliography{bc}

\end{document}